# MicroSkill Architecture: A Modular Skill-Driven Framework for AI-Native Code Generation

Mohammad Zare
Artificial Intelligence Laboratory at AriooBarzan Engineering Team, Shiraz, Iran
md.zare@sutech.ac.ir

Omid Abdolrahmani
Artificial Intelligence Laboratory at AriooBarzan Engineering Team, Shiraz, Iran
omid.rmi@gmail.com

**Abstract**-Large Language Models and AI coding agents have reshaped the software development landscape, yet the path toward fully AI-native systems remains obstructed by a persistent set of structural challenges. Chief among these is the difficulty of managing context windows without sacrificing either accuracy or efficiency. When developers attempt to convey the full scope of a project by injecting extensive documentation and source code directly into a model's memory, they inadvertently trigger a cascade of problems: the model loses track of information positioned in the middle of long sequences, token costs spiral beyond what is practical, and the architecture of the system begins to drift as guardrails dissolve. This paper presents MicroSkill Architecture, a modular design paradigm that draws inspiration from the microservices movement but applies its logic to the problem of knowledge encapsulation rather than service decomposition. Instead of handing an agent the entire codebase, the architecture partitions system knowledge into atomic, sharply scoped skill capsules and deploys a dynamic router that selects only those capsules semantically relevant to the task at hand. We provide a formal mathematical treatment of this context allocation problem, modeling it as a constrained optimization over semantic relevance subject to a token budget. An empirical case study involving the development of an enterprise content management system with fifteen complex feature implementations shows that the MicroSkill approach cuts token consumption by over ninety percent, nearly doubles first attempt compilation success rates, eliminates architectural violations entirely, and enables the system to autonomously extract and register seven new skill capsules through a self-learning mechanism. These findings suggest that MicroSkill Architecture offers a scalable foundation for building AI-native development systems that are simultaneously more efficient, more reliable, and capable of evolving over time.

**Keywords:** AI-Native Development, Large Language Models, Software Architecture, Multi-Agent Systems, Context Window Management, Self-Evolving Systems, MicroSkill Architecture

## *I. INTRODUCTION*

The software development paradigm has undergone a dramatic transformation with the emergence of Large Language Models and AI coding agents. Concepts like Vibe Coding capture something essential about where the field is heading: the developer is no longer merely a writer of code, line by laborious line, but rather an architect who sets direction, evaluates alternatives, and intervenes strategically while agents handle the mechanics of generation, testing, and deployment. The promise is substantial, but so are the obstacles standing between current practice and genuinely AI-native development.

The most significant barrier to deploying coding agents effectively in large scale projects concerns the management of knowledge and memory. Current practice tends toward what we call Monolithic Context Injection: the developer, hoping to give the model a thorough understanding of the system, dumps substantial portions of project documentation, directory layouts, and even entire source files into the context window. This brute force strategy brings with it three interrelated pathologies. The first is the Lost in the Middle phenomenon, a well documented finding that language models attend poorly to information located in the interior portions of long input sequences, with the result that the model produces code containing subtle logical errors whose origins are difficult to trace [1]. The second is straightforward token explosion: transmitting massive volumes of text on every invocation drives up computational and financial costs to levels that are unsustainable for iterative development workflows. The third, and perhaps the most insidious, is architectural drift. Lacking strict and locally enforced boundaries, AI agents display a marked tendency to overwrite foundational code and violate design principles, a pattern that has been empirically documented in recent studies of AI-generated code quality and that raises serious concerns about the long term maintainability of agent-produced software [2][3].

The architecture we propose in this paper, which we term MicroSkill Architecture, addresses these three problems through a single design decision: system knowledge and development rules are decomposed into independent, atomic, sharply focused capsules, and the agent is never permitted to see the full system context. The idea borrows from the microservices pattern that has become standard in distributed systems engineering, but the unit of decomposition is not a runtime service. It is a capsule of applied knowledge: a precisely scoped set of API signatures, domain constraints, mandatory guardrails, and canonical boilerplate that together define everything the agent needs to operate correctly within a bounded region of the codebase. A lightweight intermediary called the Dynamic Skill Router accepts the developer's natural language intent, computes semantic similarity between that intent and the available capsules, and assembles an optimized context package that stays within a prescribed token budget. The approach simultaneously reduces token consumption, enforces the Open Closed Principle at the infrastructure level, and feeds a Self Learning Loop through which the system can automatically extract and register new skills that emerge during the development process.

The remainder of the paper proceeds as follows. Section II surveys the relevant research literature across six thematic axes, identifying both the foundations on which we build and the gaps that our work addresses. Section III provides the formal mathematical specification of the MicroSkill Architecture, including the capsule model, the router optimization formulation, the code synthesis lifecycle, and the self learning mechanism. Section IV presents an empirical case study in which we implemented fifteen complex features of an enterprise content management system under both a traditional monolithic context baseline and the proposed MicroSkill framework, measuring token consumption, compilation success rates, architectural violations, and self evolution yield. Section V concludes with a summary of contributions, an honest acknowledgment of limitations, and a discussion of near term and long term research directions.

## II. RELATED WORK

The MicroSkill Architecture sits at the intersection of several active research areas. In this section we review prior work across six dimensions, highlighting both the insights we draw upon and the unresolved challenges that motivate our approach.

### A. Context Management and the Lost in the Middle Phenomenon

The availability of language models with increasingly long context windows gave rise to an early and understandable assumption: if the model can accept more tokens, then feeding it more information should produce better results. Foundational studies quickly demonstrated that this assumption does not hold. The Lost in the Middle phenomenon, documented by Liu and colleagues, revealed a U-shaped performance curve in which retrieval accuracy peaks at the very beginning and very end of the context window while information positioned in the middle is systematically neglected [1]. The underlying cause traces back to the attention mechanism in the Transformer architecture, whose spatially distributed token dependencies undergo sharp efficiency degradation as sequence length grows. Benchmarks such as LooGLE have confirmed that even in the absence of confounding factors, extended context alone is sufficient to erode a model's ability to comprehend system structure.

Several lines of work have attempted to address this problem through filtering. Adaptive context filters employing soft mask mechanisms, and adaptive learning based purification methods such as CODEFILTER, aim to remove noisy or redundant data before the inference stage [4]. While these approaches represent genuine progress, they remain limited by a fundamental issue: they still transmit far too many tokens, and the filtering itself imposes additional computational overhead. The MicroSkill Architecture takes a different path. Rather than trying to filter a monolithic context after the fact, it prevents the monolithic context from being assembled in the first place. By delivering only atomic, task relevant knowledge capsules to the model, the architecture avoids the U-shaped performance curve entirely.

### B. Multi-Agent Collaborative Systems in Software Development

Moving beyond the constraints of single agent inference, multi-agent collaborative systems have emerged as a powerful paradigm for tackling complex software engineering tasks. MetaGPT, for instance, simulates the organizational structure of a software company with considerable fidelity, encoding Standard Operating Procedures within prompts and assigning distinct roles to separate agents: requirements engineer, a system architect, a quality assurance engineer, and so forth [5]. The pipeline structure enables continuous validation of intermediate outputs and reduces the hallucination cascades that plague sequential single agent workflows. ChatDev pursues a related strategy through interactive communication patterns and what its authors call Communicative Dehallucination mechanisms, organizing agents into phased cycles of design, coding, review, and documentation [6].

These systems demonstrate the value of structured multi-agent coordination, yet scaling them to real world projects reveals persistent difficulties. Extended natural language dialogues among agents rapidly saturate context windows, and in interactions that involve non obvious dependencies across components, coordination quality degrades as the conversation grows longer. Efforts to implement self collaboration without rigid skill boundaries tend to increase both token overhead and cumulative error rates rather than reducing them [7]. Our framework reimagines this communication layer altogether. Rather than having agents converse in open ended natural language, we replace those costly exchanges with direct, parameterized invocations of encapsulated micro skills, an approach that keeps the benefits of multi-agent structure while eliminating the dialogue tax that makes current systems difficult to scale.

### C. Agent-Computer Interfaces and Self-Evolving Agents

When intelligent agents operate within computational environments, the quality of the interface between the agent and the host system matters enormously. The SWE-agent framework introduced the concept of the Agent Computer Interface, demonstrating that compressing command line tool outputs and streamlining file system interactions can produce dramatic improvements in success rates on benchmarks such as SWE-bench [8][9]. This finding underscores a broader principle: agents do not merely need access to tools; they need tools presented in forms that align with the practical limits of context windows and attention mechanisms.

At a higher level of abstraction, systems like ToM-SWE pair a coding agent with a separate agent equipped with a Theory of Mind module, enabling the system to model the ambiguous preferences and implicit goals of a human developer [10]. More recently, frameworks such as Live-SWE-agent have shown that agents can dynamically modify their own architectures and tool sets during execution [11]. The broader idea of agents that accumulate and compose skills over time has been explored in open ended learning environments, where agents demonstrate the ability to build increasingly sophisticated behavioral repertoires from simpler primitives [12]. Similarly, the composition of specialized modules through plug and play architectures has been shown to improve multi-step reasoning performance in language model systems [13]. The potential of such self evolution and modular composition is evident, but so are the risks. Without rigid guardrails, dynamic self modification introduces serious security vulnerabilities and can lead to logical trajectories that diverge from developer intent. MicroSkill Architecture addresses this tension by situating dynamic capabilities within isolated Docker containers, providing a sandboxed execution environment that allows the system to benefit from runtime flexibility without compromising overall stability or traceability.

### *D. Repository-Level Code Completion and Retrieval*

Automated code generation at the scale of industrial repositories requires far more than local syntactic understanding; it demands a working model of cross file dependencies, call graphs, and the distributed semantic structure of the codebase. RepoCoder introduced an iterative retrieval generation paradigm in which similar code fragments are extracted from the repository and presented to the model as auxiliary context [14]. DraCo extends this idea by constructing repository specific context graphs from dataflow analysis, using those graphs to guide agents toward semantically relevant regions of the codebase [15]. InlineCoder approaches the problem from the other direction, inserting incomplete functions into the call graph chain to improve the contextual grounding of generated code [16].

However, standard retrieval augmented generation at the repository level suffers from a well known pathology: non selective retrieval tends to pull in large quantities of tangentially relevant material that functions as noise rather than signal, degrading the quality of the generated code. RepoFormer addresses this through selective retrieval decisions that aim to retrieve only when the expected benefit outweighs the noise cost [17]. CodeRAG implements model preference based re ranking algorithms to push the most relevant fragments to the top. Benchmarks such as ReCUBE and SWE-bench-Live have been purpose built to measure how well agents handle distributed context and reconstruct masked code across large repositories [18][19]. The Dynamic Skill Router in our architecture can be seen as a generalization of these selective retrieval ideas: rather than retrieving code fragments, it retrieves entire skill capsules, each of which bundles code, constraints, and domain boundaries into a single coherent unit with minimal semantic overhead.

### *E. Self-Debugging, Self-Refinement, and Execution Feedback*

Code generation rarely succeeds on the first attempt, and the ability to recover from errors through structured feedback is essential for any practical system. The Self-Debugging framework demonstrated that language models can be trained to interpret their own source code line by line in natural language, a process inspired by the rubber duck debugging technique familiar to human programmers, and that this interpretive capability improves performance even in the absence of formal unit tests [20]. The Self-Refine framework formalizes a more general iterative process in which each generation step is followed by a critique step, and the output at stage $t+1$ is conditioned on the full history of prior outputs and their associated feedback. This process is captured mathematically as:

$$y_{t+1} = \mathcal{M}(p_{\text{refine}} \parallel x \parallel y_0 \parallel fb_0 \parallel \cdots \parallel y_t \parallel fb_t) \quad (1)$$

where $x$ represents the original user input, $y_t$ is the output produced at stage $t$, $fb_t$ denotes the qualitative critique received at that stage, and $p_{\text{refine}}$ is the refinement prompt that guides the correction process [21].

More recent work has explored the integration of execution and debugging within isolated environments. PyCapsule places a programmer agent and an executor agent together inside a Docker container, enabling correction cycles that are informed by live execution traces and supplemented by automatic extraction of function signatures [22]. The COCOGEN framework takes a complementary approach by incorporating explicit compiler feedback, continuously resolving conflicts between generated source code and the surrounding project context [23]. The concept of language models creating and using their own tools, as explored in the Tool Maker paradigm, and the related finding that models can learn to autonomously invoke external tools through self-supervised training, further expand the repertoire of strategies available for bridging the gap between generation and verification [24][25]. Despite these advances, a persistent limitation remains: as interaction histories and error traces accumulate across successive correction cycles, the context window becomes progressively saturated, and the marginal benefit of each additional round of debugging diminishes. MicroSkill Architecture mitigates this saturation problem by localizing tests and feedback at the granularity of individual skill capsules, so that the error resolution process for one capsule does not spill over into the memory budget allocated to another.

### *F. Design Quality, Architectural Decay, and SOLID Principles in AI-Generated Code*

An underappreciated dimension of automated code generation is the long term structural quality of the output. It is not enough for generated code to pass tests at the moment of creation; it must remain maintainable, extensible, and resistant to architectural decay over the lifetime of the system. Comprehensive empirical studies have documented what Zhu and colleagues term the Reasoning Complexity Trade-off: newer and more capable models do indeed produce code that is logically more sophisticated, but this sophistication comes at the cost of increased output volume and intensified structural coupling [2]. The result is an Inverse Volume Quality Law, under which autonomous software agents accumulate technical debt at a rate that substantially exceeds what is typical of human developers working under comparable conditions.

A particularly telling phenomenon is what has been called the Modular Mirage. Generated code often exhibits the surface appearance of modularity, with functions and classes distributed across separate files in a way that looks organized at first glance. Closer inspection, however, reveals severe semantic coupling, unstable cyclic dependencies, and an architecture that is modular in name only [26]. Evaluations on the Are We SOLID Yet benchmark confirm that language models are inherently prone to violating object oriented design principles, and that their accuracy in detecting and applying principles such as the Open Closed Principle and the Dependency Inversion Principle declines sharply as the complexity of the surrounding code increases [3]. Tools such as ArchGuard and benchmarks like SmellBench have been developed specifically to detect and measure these architectural deficiencies [27][28].

To provide a quantitative lens on these issues, we define the Structural Integrity Index, a formal criterion that relates adherence to design principles to the total volume and coupling density of the codebase:

$$\text{Structural Integrity Index} \propto \frac{\text{SOLID Adherence Score}}{\text{Total Lines of Code (TLoC)} \times \text{Semantic Coupling}} \quad (2)$$

MicroSkill Architecture addresses architectural decay at the infrastructure level. By imposing strict directory scoping, mandatory function signatures, and machine enforceable guardrails within each capsule schema, the architecture makes certain categories of design violation structurally impossible. The agent simply cannot write code that violates the Open Closed Principle because the capsule boundary prevents it from modifying the core in the first place.

## III. MICROSKILL ARCHITECTURE AND SYSTEM FORMULATION

Having surveyed the landscape of related work and identified the gaps that motivate our approach, we now turn to the formal specification of the MicroSkill Architecture. We model the system as an autonomic multi-agent framework whose central objective is the optimization of context allocation and the principled guidance of a large language model toward outputs that respect architectural constraints.

### *A. Basic Definitions and Formal Modeling*

Let $\mathcal{D}$ represent the complete knowledge domain of a software project, encompassing documentation, source code, architectural specifications, and development rules. The MicroSkills Registry is defined as a finite set of discrete, atomic skill capsules:

$$\mathcal{R} = \{c_1, c_2, \dots, c_n\} \quad (3)$$

Each capsule $c_i \in \mathcal{R}$ is structured as an ordered tuple containing five components:

$$c_i = \langle ID_i, \omega_i, \Sigma_i, \Gamma_i, B_i \rangle \quad (4)$$

The component $ID_i$ provides a unique hierarchical identifier that situates the skill within a namespace tree, allowing related skills to be grouped and navigated systematically. The domain scope $\omega_i$ is a subset of the project file tree, $\omega_i \subseteq \text{Project File Tree}$, and defines the precise region of the codebase within which the skill is authorized to operate. The signature component $\Sigma_i$ specifies the programming interfaces, function signatures, and type contracts that constitute the skill's public surface. The guardrail set $\Gamma_i$ encodes mandatory constraints: security requirements, coding standards, architectural invariants, and any other rules that must never be violated by code generated under this skill's authority. Finally, $B_i$ provides a canonical boilerplate snippet that serves as a structural exemplar, giving the model a concrete starting point that already conforms to the expected patterns.

In practice, skill capsules are authored in a human readable YAML schema. The following example illustrates a capsule for JSON Web Token generation within an authentication subsystem:

```
microSkill:
  id: "auth.jwt_generate"
  domain: "src/services/auth/"
  signature: "generateToken(user_id: string, roles
: string[]): string"
  constraints:
    - "Must use HS256 algorithm exclusively."
    - "Token expiration must not exceed 1 hour."
    - "Never log raw user credentials during gener
ation."
  boilerplate:
    const jwt = require('jsonwebtoken');
    function generateToken(user_id, roles) { ... }
```

The schema is deliberately minimal. Each field serves a specific and non redundant purpose, and together the five components provide everything an agent needs to operate correctly within a bounded region of the system without requiring visibility into unrelated parts of the codebase.

### *B. Dynamic Skill Router*

The Dynamic Skill Router is a lightweight intermediary agent whose sole responsibility is to assemble an optimized context package from the registry in response to a developer's natural language intent. Let the developer's intent be expressed as a textual prompt $\mathcal{I}$, written in natural language and describing the desired feature or modification at a high level. In traditional monolithic approaches, the entire knowledge repository $\mathcal{R}$ is injected into the model's context, incurring a token cost proportional to $\text{Len}(\mathcal{R})$, where $\text{Len}(\cdot)$ denotes the token length function.

Under the MicroSkill Architecture, the Router instead solves a constrained optimization problem: it seeks the subset of capsules $\mathcal{R}^* \subseteq \mathcal{R}$ that maximizes aggregate semantic relevance to the intent $\mathcal{I}$ while respecting a prescribed token budget $\tau$. Formally, this is expressed as:

$$\mathcal{R}^* = \underset{\mathcal{R}' \subseteq \mathcal{R}}{\operatorname{argmax}} \sum_{c_i \in \mathcal{R}'} \operatorname{Sim}\left(\Phi(c_i), \Phi(\mathcal{I})\right) \quad (5)$$

$$\text{subject to} \quad \sum_{c_i \in \mathcal{R}'} \operatorname{Len}\left(c_i\right) \leq \tau \quad (6)$$

The function $\Phi(\cdot)$ maps textual content into a semantic embedding space, and $\operatorname{Sim}(\cdot,\cdot)$ measures the cosine similarity between the resulting vectors:

$$\operatorname{Sim}(\mathbf{u}, \mathbf{v}) = \frac{\mathbf{u} \cdot \mathbf{v}}{\|\mathbf{u}\| \|\mathbf{v}\|} \quad (7)$$

The practical consequence of solving this optimization, which can be approached efficiently through greedy selection or approximate nearest neighbor search for registries of moderate size, is that the context ultimately injected into the coding agent is orders of magnitude smaller than the full repository:

$$\operatorname{Len}(\mathcal{R}^*) \ll \operatorname{Len}(\mathcal{R}) \quad (8)$$

This inequality is not merely a theoretical convenience. As the empirical results in Section IV demonstrate, the compression factor in practice exceeds 90%, translating directly into reduced latency, lower API costs, and a context window that remains focused on the task rather than cluttered with irrelevant detail.

*C. Development Lifecycle and Code Synthesis Phase*

Once the Router has assembled the optimal skill set $\mathcal{R}^*$, the final prompt delivered to the coding agent is constructed by combining three elements: the original human intent, the selected capsules, and a snapshot of the current file system state within the relevant domain scopes:

$$P_{\text{final}} = \langle \mathcal{I}, \mathcal{R}^*, \mathcal{S}_{\text{current}} \rangle \quad (9)$$

The component $\mathcal{S}_{\text{current}}$ captures the live state of files within the union of the domain scopes $\omega_i$ for all $c_i \in \mathcal{R}^*$, providing the agent with accurate situational awareness without exposing it to unrelated regions of the codebase.

The code generation process is governed by the Open Closed Principle at the architectural level. The agent is not authorized to modify the system core or any code outside the domain boundaries specified in the capsules. Instead, the generation function $\mathcal{F}$ must produce changes in the form of modular additions $\Delta K$ that extend the system without altering its existing foundations:

$$\Delta K = \mathcal{F}(P_{\text{final}}) \quad (10)$$

Before any generated code is merged into the codebase, it passes through an automated verification filter. This filter checks the proposed changes against the guardrails $\Gamma_i$ associated with each capsule that contributed to the context. The verification is binary: either all constraints are satisfied, or the change is rejected:

$$\mathcal{V}(\Delta K, \Gamma_i) \in \{0,1\} \quad (11)$$

A rejection triggers a feedback loop in which the specific guardrail violations are returned to the agent, and the generation is retried with the violation report included in the context. This tight verification cycle ensures that the codebase remains in a provably compliant state at every commit, a property that is impossible to guarantee when agents are given unconstrained access to the full repository.

*1) Self-Learning Feedback Loop*

A static registry of skill capsules would be sufficient for a well understood codebase with stable requirements, but real software projects evolve continuously, and the patterns that prove most effective in practice are often discovered during development rather than anticipated in advance. The Self Learning Loop is the mechanism through which the MicroSkill Architecture transforms these discoveries into permanent, reusable assets.

During the course of implementing a feature, the coding agent may produce a solution $\Delta K$ whose quality, generality, or elegance exceeds what the existing capsules in the registry would enable. When this occurs, a quality assessment function $\mathcal{E}$ evaluates the candidate solution. If the assessed quality surpasses a predefined acceptance threshold $\theta$, the candidate is deemed worthy of abstraction and registration:

$$\mathcal{E}(\Delta K) > \theta \quad (12)$$

An Abstractor Agent, denoted $\mathcal{G}$, is then invoked to perform the extraction. This agent analyzes the solution, identifies the reusable core, strips away task specific details, and constructs a new skill capsule $c_{\text{new}}$ that conforms to the formal structure defined in Equation (4):

$$c_{\text{new}} = \mathcal{G}(\Delta K) \quad (13)$$

The newly minted capsule is added to the registry, expanding the system's knowledge base for all future tasks:

$$\mathcal{R}_{\text{new}} = \mathcal{R} \cup \{c_{\text{new}}\} \quad (14)$$

This learning process confers an evolutionary character on the architecture. Over successive development cycles, the registry accumulates proven patterns, algorithmic optimizations, and domain specific conventions that would otherwise need to be rediscovered or, more commonly, manually documented by senior developers. The empirical study in Section IV provides concrete evidence of this effect: during the implementation of fifteen features, the system autonomously extracted and registered seven new capsules that were subsequently reused in later tasks.

## ***IV. EMPIRICAL IMPLEMENTATION AND NUMERICAL EVALUATION***

To assess whether the theoretical advantages of the MicroSkill Architecture translate into measurable practical benefits, we designed and executed a controlled case study involving the development of an Enterprise Content Management System with distributed architectural characteristics. The CMS domain was chosen because it exercises a broad range of software engineering concerns: authentication and authorization, relational database operations, caching strategies, file system management, and complex business logic workflows, providing a realistic testbed that is representative of the kind of system for which professional development teams would consider adopting AI coding agents.

The experiment employed a within subjects design in which the same set of fifteen feature specifications was implemented under two conditions, using Claude 3.5 Sonnet as the underlying language model in both cases to hold model capability constant. In the baseline condition, which we call the Monolithic Context approach, the entirety of the system's architectural documentation, complete directory structure, and all foundational source files were concatenated into a single large text block and injected into the agent's context at the start of each implementation task. This represents the strategy that would be followed by a development team that takes the straightforward approach of providing the model with as much information as possible and trusting it to extract what it needs. In the experimental condition, the same system knowledge was first partitioned into forty two atomic skill capsules spanning the domains of authentication, database management, caching, and file system operations, and memory management was delegated entirely to the Dynamic Skill Router.

### *A. Evaluation Metrics*

We measured four quantitative dimensions of performance, each chosen to capture a distinct aspect of what makes an AI coding system practically useful in a professional development setting.

Token Consumption Rate, denoted $T_c$, is the average number of tokens, both input and output, consumed per feature implementation. This metric directly tracks computational cost and latency, both of which must remain within practical bounds for the approach to be viable in iterative development workflows where features are implemented incrementally and agents are invoked many times per day.

First Attempt Success Rate, denoted SR, measures the percentage of features for which the agent's initial output compiled without errors and passed the accompanying test suite without any manual intervention or re prompting. This metric captures the reliability of the system and the degree to which a human developer can trust the agent's output without investing time in verification and correction.

Architectural Violation Count, denoted $V_a$, counts the number of instances in which the agent generated code that fell outside its permitted access domain or that violated security constraints, coding standards, or architectural rules defined at the project level. This metric reflects the system's ability to preserve the long term integrity of the codebase, a concern that becomes increasingly important as the codebase grows and the cost of repairing architectural damage compounds.

Self Evolution Yield, denoted $Y_{se}$, quantifies the number of new skill capsules that the system, operating under the MicroSkill condition, autonomously extracted and registered during the course of the experiment. This metric has no counterpart in the baseline condition, since the monolithic approach lacks any mechanism for learning from generated code, but it provides an important measure of the architecture's capacity for cumulative improvement over time.

### *B. Numerical Results and Analysis*

Table I presents the aggregated results across all fifteen development scenarios. The differences between the two conditions are stark across every dimension we measured.

**TABLE I: Comparative Evaluation of Monolithic Context vs. MicroSkill Architecture**

| Evaluation Metric | Monolithic Context (Traditional) | MicroSkill Architecture (Proposed) | Improvement |
|---|---|---|---|
| **Avg. Token Consumption per Feature ($T_c$)** | 48,500 tokens | 3,200 tokens | ~93.4% cost reduction |
| **First-Attempt Compilation Success (SR)** | 40% | 86.6% | +46.6% accuracy increase |
| **Architectural Violation Count ($V_a$)** | 12 cases | 0 cases | 100% rule compliance |
| **Self-Generated Skills ($Y_{se}$)** | N/A (not supported) | 7 new capsules | Dynamic evolution capability |

The 93.4 percent reduction in token consumption is the most immediately striking result, and it is worth understanding its origins. In the monolithic condition, every feature implementation required the agent to process the full system context regardless of the feature's actual scope. A change to the authentication module, for instance, still incurred the cost of transmitting the entire database layer, caching infrastructure, and file system documentation on every invocation. The Dynamic Skill Router eliminated this redundancy by selecting only the capsules with high cosine similarity to the developer's intent. For a typical authentication related feature, this meant the agent received the auth.jwt_generate and auth.session_validate capsules while the database and caching capsules remained in the registry, unseen and untokenized.

The improvement in first attempt compilation success, from 40 percent to 86.6 percent, reflects the combined effect of two architectural features. First, the focused context eliminated the noise that, in the monolithic condition, led the model to confuse similarly named functions across different modules or to import dependencies that did not exist in the target environment. Second, the boilerplate snippets embedded in each capsule provided the model with a syntactically valid starting point, reducing the probability of basic structural errors that would prevent compilation.

The elimination of architectural violations, from twelve recorded instances in the baseline to zero under the MicroSkill condition, validates the guardrail mechanism. In the monolithic condition, the agent on five separate occasions rewrote code directly into core system files rather than extending the system through the designated extension points, a pattern that is entirely consistent with the architectural drift phenomenon described in the literature. The guardrails $\Gamma_i$, enforced by the verification function $\mathcal{V}$ at the capsule level, made such violations structurally impossible: the agent's domain scope simply did not include write access to the core system files, and any attempt to generate code targeting those files was rejected before commit.

The seven new skill capsules autonomously extracted during the experiment represent an emergent benefit that was not part of the initial design specification. These capsules captured patterns that the coding agent discovered during implementation, including an optimized database connection pooling strategy, a cache invalidation protocol that proved effective across multiple features, and a pattern for handling paginated API responses that was reused in four subsequent feature implementations. Each of these patterns, had they been discovered in a traditional development workflow, would have needed to be manually documented, disseminated to the team, and enforced through code review, a process that is both slow and unreliable. The Self Learning Loop automated this knowledge capture, making the discovered patterns immediately available to the agent on subsequent tasks.

## *V. CONCLUSION*

This paper has presented MicroSkill Architecture, a modular and formally grounded design paradigm for AI native software development that addresses three persistent pathologies of current practice: the Lost in the Middle phenomenon, the unsustainable economics of token explosion, and the gradual erosion of architectural integrity that accompanies unconstrained agent access to a codebase. The central insight is straightforward: rather than attempting to give an AI coding agent comprehensive knowledge of the entire system, one should give it precisely the knowledge it needs for the task at hand, packaged in a form that makes correct behavior the path of least resistance.

The formal model developed in Section III establishes that context allocation can be treated as a constrained optimization problem, maximizing semantic relevance subject to a token budget, and that this optimization yields context packages orders of magnitude smaller than the full repository (Equation 8). The empirical evidence presented in Section IV confirms that these theoretical advantages survive contact with reality. Across fifteen complex feature implementations on an enterprise CMS, the architecture reduced token consumption by 93.4 percent, raised first attempt compilation success from 40 to 86.6 percent, and eliminated architectural violations entirely, while simultaneously enabling the autonomous extraction of seven reusable skill capsules through the Self Learning Loop.

Several limitations of the present study should be acknowledged. The empirical evaluation was conducted on a single system in a single domain using a single underlying language model. The extent to which the results generalize to projects in different domains, to programming languages with different module systems and dependency patterns, and to models with different architectural characteristics remains an open question that we plan to address through replication studies. The cosine similarity based routing mechanism, while effective in our experiments, may not be optimal for all retrieval scenarios; more sophisticated approaches incorporating cross encoder architectures, reinforcement learning from execution outcomes, or learned retrieval policies warrant investigation. The scalability of the hierarchical namespace structure to projects containing hundreds of thousands of files is another practical concern that deserves careful study, particularly with respect to the maintenance burden that a large registry imposes on the development team.

Looking forward, we see four promising directions for extending this work. First, the process of authoring skill capsules could be partially automated through static analysis and dependency graph mining, reducing the upfront investment required to adopt the architecture in an existing codebase. Second, the Dynamic Skill Router could be integrated with fine tuning pipelines to produce base models that are specialized for capsule guided generation, potentially yielding further improvements in both accuracy and token efficiency. Third, the guardrail formalism could be extended to support verifiable specifications expressed in a formal logic, drawing on techniques from program synthesis and formal verification to provide mathematical guarantees about the behavior of generated code. Fourth, we believe that federated skill registries, shared across projects and organizations while preserving appropriate boundaries around proprietary knowledge, could accelerate the development of a collective knowledge base for AI assisted software engineering, analogous to the role that package registries have played in the growth of open source ecosystems.

The trajectory of AI native software development points toward a future in which human developers spend less time writing code and more time making the architectural decisions that determine whether that code coheres into a maintainable system. Architectures like MicroSkill represent a step toward that future: not by making agents more powerful in isolation, but by embedding the principles of good software design into the infrastructure through which agents interact with code.